# Synthesis and Properties of La$_{1-x}$Sr$_x$NiO$_3$ and La$_{1-x}$Sr$_x$NiO$_2$

Mengwu Huo, Zengjia Liu, Hualei Sun, Lisi Li, Hui Lui, Chaoxin Huang, Feixiang Liang, Bing Shen, and Meng Wang*

*Center for Neutron Science and Technology, School of Physics, Sun Yat-Sen University, Guangzhou, 510275, China*

Abstract: Superconductivity has been realized in films of La$_{1-x}$Sr$_x$NiO$_2$. Here we report synthesis and characterization of polycrystalline samples of La$_{1-x}$Sr$_x$NiO$_3$ and La$_{1-x}$Sr$_x$NiO$_2$ ($0 \leq x \leq 0.2$). Magnetization and resistivity measurements reveal that La$_{1-x}$Sr$_x$NiO$_3$ are paramagnetic metal and La$_{1-x}$Sr$_x$NiO$_2$ exhibit an insulating behavior. Superconductivity is not detected in bulk samples of La$_{1-x}$Sr$_x$NiO$_2$. The absence of superconductivity in bulk La$_{1-x}$Sr$_x$NiO$_2$ may be due to the generation of hydroxide during reduction or a small amount of nickel impurities. The effect of interface in films of La$_{1-x}$Sr$_x$NiO$_2$ may also play a role for superconductivity.

The discovery of superconductivity in film samples of Nd$_{0.8}$Sr$_{0.2}$NiO$_2$ has drawn much attention on nickel oxide materials[1]. In fact, LaNiO$_3$ was studied in 1983 before the discovery of superconductivity in copper oxide materials[2]. La$_{2-x}$Sr$_x$NiO$_4$ were investigated because of the structural similarity to the high temperature superconducting (SC) cuprate perovskites La$_{2-x}$Sr$_x$CuO$_4$. Superconductivity has also been predicted in nickelate heterostructures like LaNiO$_3$/La$M$O$_3$ ($M$ is a trivalent cation, like Al or Ga) superlattices[3, 4], while superconductivity has not been realized in nickel oxide materials with the oxidation states of Ni$^{3+}$ and Ni$^{2+}$. It was suggested that superconductivity may exist in low spin oxidation state of Ni$^{1+}$ with a square coordination with O$^{2-}$ ions[5-7]. However, the low valence state of nickel is difficult to stabilize through conventional high-temperature-solid state reaction synthesis method.

Tuning the valence states of Ni ions in nickelate thin film heterostructures by combining CaH$_2$ reductant and carrier doping greatly stimulated scientists to explore superconductivity and other novel physical properties in nickel-based oxides[1, 8-17]. Infinite layer $R$NiO$_2$ ($R$ = Nd, La, and Pr) is isostructural to CaCuO$_2$ with two-dimensional NiO$_2$ planes stacking as NiO$_2$-$R$-NiO$_2$-$R$ layers as shown in Fig. 1. CaCuO$_2$ is one of the parent compounds of high-temperature superconductors. Through hole-doping the antiferromagnetic (AF) order can be suppressed and superconductivity emerges in (Ca$_{1-x}$Sr$_x$)$_{1-y}$CuO$_{2-\delta}$ with the maximum SC transition temperature of $T_C$ = 110 K[18, 19]. Furthermore, the spin configuration of Ni$^{1+}$ 3$d^9$ is similar to that of Cu$^{2+}$ in curates. SC domes have been observed in films of $R_{1-x}$Sr$_x$NiO$_2$ ($R$ = La, Nd, and Pr)[9-11, 20], La$_{1-x}$Ca$_x$NiO$_2$[21] and Nd$_6$Ni$_5$O$_{12}$[13]. While the superconductivity observed in nickel-based oxides are all in the form of thin film samples, it is worthwhile to explore superconductivity in bulk samples. Recently bulk samples $R_{1-x}$Sr$_x$NiO$_2$($R$ = Sm, Nd) [22-24] and La$_{1-x}$Ca$_x$NiO$_2$[25] were synthesized whereas superconductivity was not detected. So far, electrical and magnetic properties of Sr doped bulk samples of La$_{1-x}$Sr$_x$NiO$_2$ have not been reported, where superconductivity has been observed in film samples.

In this paper, we report synthesis and systematic studies of the bulk samples of La$_{1-}$

$_x$Sr$_x$NiO$_3$ and La$_{1-x}$Sr$_x$NiO$_2$ (0 ≤ x ≤ 0.2). La$_{1-x}$Sr$_x$NiO$_3$ are synthesized using molten KOH as flux. La$_{1-x}$Sr$_x$NiO$_2$ are obtained via a topochemical reduction process from La$_{1-x}$Sr$_x$NiO$_3$ using CaH$_2$ as a reducing agent. Measurements of resistivity, magnetization, and heat capacity were carried out to explore the magnetic and electrical properties. Results of electronic transport show that La$_{1-x}$Sr$_x$NiO$_3$ (0 ≤ x ≤ 0.2) are good metals, while La$_{1-x}$Sr$_x$NiO$_2$ exhibit insulating behavior and no superconductivity is observed.

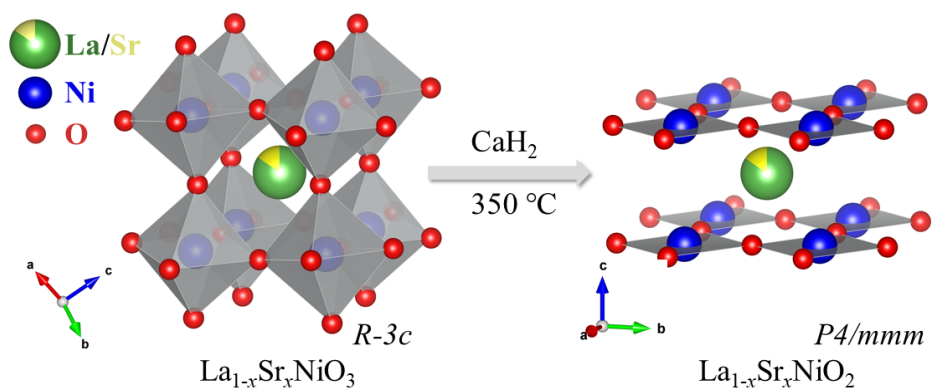

**Figure 1.** Structures of La$_{1−x}$Sr$_x$NiO$_3$ (left) and La$_{1−x}$Sr$_x$NiO$_2$ (right) (0 ≤ x ≤ 0.2). Upon low temperature reduction with CaH$_2$ as a reducing agent, the powder samples undergo a topotactic transition from the perovskite phase to the infinite-layer phase.

La$_{1-x}$Sr$_x$NiO$_3$ (0 ≤ x ≤ 0.2) were synthesized using KOH as the flux[26]. La$_2$O$_3$ was preheated at 1100 °C for one day to remove water then weighed in a glove box filled with argon. Stoichiometric ratios of powders of NiO (99.99%), La$_2$O$_3$ (99.9%), and SrCO$_3$ (99.99%) were mixed and heated at 1200 °C for one day. The flux of KOH was heated at 450 °C for 12 h, followed by the mixture of the nickelate oxide powders and dwelled at 450 °C for 10 h. KOH was removed from powders of La$_{1-x}$Sr$_x$NiO$_3$ by deionizing water. Thereafter the powders were dried at 120 °C. Finally, dark powders of La$_{1-x}$Sr$_x$NiO$_3$ were obtained. The topotactic reduction process was demonstrated in Fig. 1. The process was as follows: La$_{1-x}$Sr$_x$NiO$_3$ were placed in a crucible and CaH$_2$ was wrapped with an aluminum foil. They were sealed in an evacuated silica tube as a molar ratio of 1 to 4. La$_{1-x}$Sr$_x$NiO$_2$ were obtained after heating the mixture at 350 °C for 12 h.

Crystal structures of the samples were investigated by X-ray diffraction (XRD) at 300 K. The diffraction data were refined by the Rietveld method[27]. Energy dispersive X-ray spectroscopy (EDS, EVO, Zeiss) were employed to determine the compositions of the powder samples. Magnetization and resistivity in the temperature range of 3 ~ 300 K and heat capacity measurements in the range of 3 ~ 200 K were conducted on a physical property measurement system (PPMS, Quantum Design). The electrical resistivity was measured on a pressed bar sample using a standard 4-prob technique.

The crystal structures of La$_{1-x}$Sr$_x$NiO$_3$ and La$_{1-x}$Sr$_x$NiO$_2$ are shown in Fig. 1. XRD patterns of the two compounds are depicted in Fig. 2(a) and (b), respectively. The XRD patterns of LaNiO$_3$ can be indexed as a rhombohedral unit cell with the lattice constants $a = b = 5.4559$ (2), $c = 13.1359$ (7) Å, and LaNiO$_2$ as a tetragonal unit cell with the lattice constants $a = b = 3.9544(2)$, $c = 3.3814(3)$ Å. The results are consistent with that of reported previously[28-31]. The XRD patterns for Sr doped compounds indicate that the powder samples were a single phase. Variation of the lattice parameters as a function of the Sr contents is shown in Fig. 2(c). The lattice constants $a$ and $b$ decrease with the doping level $x$ while the distance between layers $c$ increases. The trend of the lattice constant variation with doping is consistent with that observed in $Nd_{1-x}Sr_xNiO_3$, revealing that the Sr doping is successful in $La_{1-x}Sr_xNiO_3$ and $La_{1-x}Sr_xNiO_2$ [24, 32]. The compositions of $La_{1-x}Sr_xNiO_2$ measured by EDS as shown in Table 1 also suggest that the Sr substitution for La is successful.

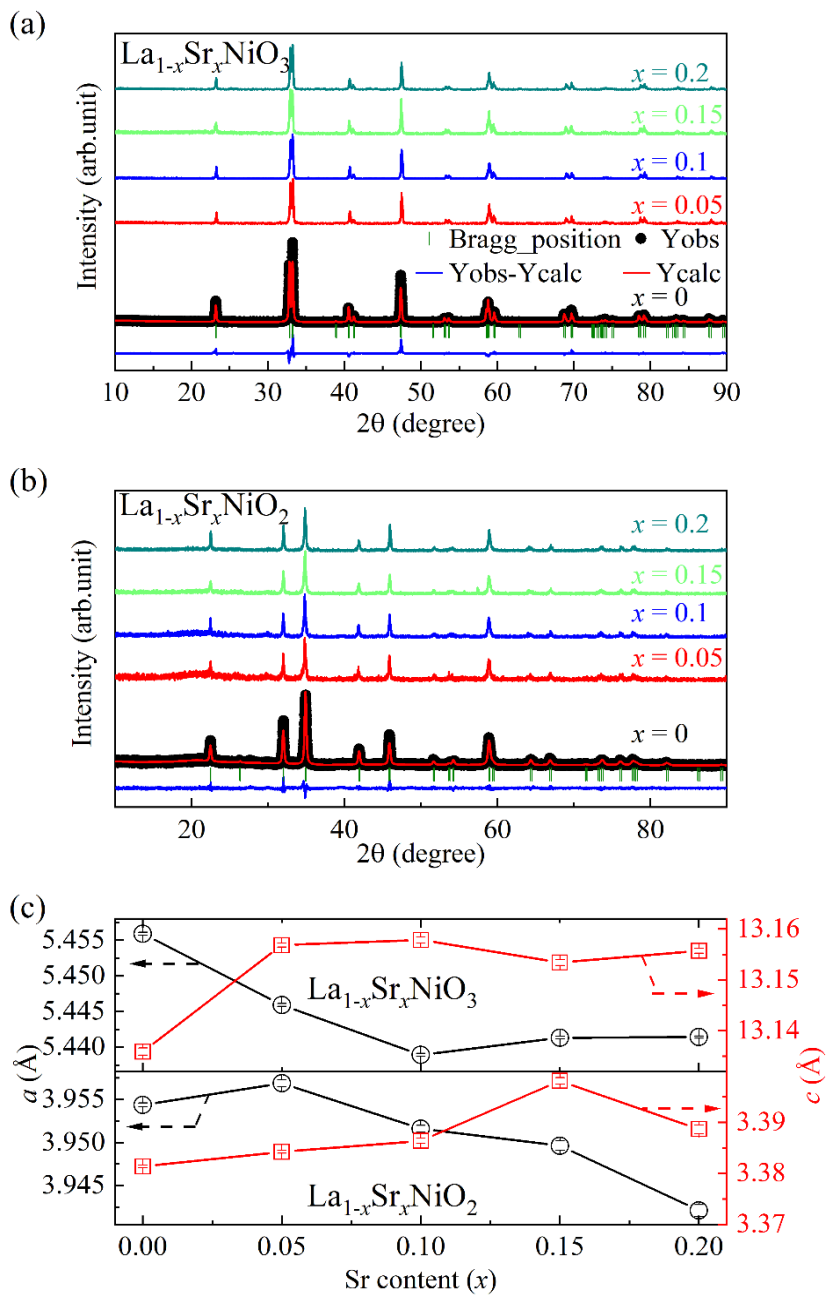

**Figure 2.** Ambient powder XRD patterns of (a) $La_{1-x}Sr_xNiO_3$ and (b) $La_{1-x}Sr_xNiO_2$ ($0 \leq x \leq 0.2$) with corresponding Rietveld fitting results by using the *R-3c* space group and *P4/mmm*, respectively.

The black dots represent the observed data, the red lines represent the fitting curves, and the blue lines illustrate the differences between the data and fitting. The olive short vertical lines mark the Bragg positions. For comparison the data are shifted vertically. (c) Lattice parameters of La$_{1-x}$Sr$_x$NiO$_3$ (top) and La$_{1-x}$Sr$_x$NiO$_2$ (bottom) ($0 \leq x \leq 0.2$) as a function of the content of Sr. The error bars are obtained from our Rietveld refinement using the Fullprof software.

Table 1. Results of EDS on La$_{1-x}$Sr$_x$NiO$_2$ ($0 \leq x \leq 0.2$).

| Nominal $x$ | La | Sr | Ni |
| --- | --- | --- | --- |
| 0 | 0.90(14) | - | 1.00(8) |
| 0.05 | 1.00(7) | 0.04(1) | 1.00(5) |
| 0.10 | 0.90(20) | 0.06(1) | 1.00(13) |
| 0.15 | 0.80(7) | 0.10(2) | 1.00(4) |
| 0.20 | 0.79(12) | 0.18(2) | 1.00(6) |

Figure 3 displays temperature dependence of the physical properties of La$_{1-x}$Sr$_x$NiO$_3$ ($0 \leq x \leq 0.2$). The resistivity and magnetization of LaNiO$_3$ are consistent with previous reports that reveal LaNiO$_3$ exhibiting a metallic state with large electronic correlations[33, 34], as shown in Fig. 3(a) and (b). The upturn in resistivity below 50 K has been observed in bulk LaNiO$_{3-\delta}$ and suggested to result from defection of oxygen during the synthesis process[35, 36]. Upon Sr doping, the metallic behavior of La$_{1-x}$Sr$_x$NiO$_3$ is remained, while the magnitude of resistivity changes.

The magnetization of LaNiO$_3$ reveals a paramagnetic behavior from 3 to 300 K. Kinks are identified at ~22 K in doped compounds. The kink is obvious for La$_{0.8}$Sr$_{0.2}$NiO$_3$ measured with $\mu_0H = 0.1$ T and zero-field cooled (ZFC) mode as shown in Fig. 3(b). The temperatures of the kinks for different compounds locate at the same temperature. To ascertain the origin of the kinks, we carried out measurements of magnetization and specific heat as a function of temperature at various magnetic fields for La$_{0.8}$Sr$_{0.2}$NiO$_3$. The magnetization is gradually suppressed as the magnetic field increasing, While the temperature of the anomaly is barely changed [Fig. 3(c)]. We repeated the process of synthesis and measurements. The kinks in magnetization for doped samples are repeatable. However, an anomaly around 22 K is not observed on specific heat at 0 and 5 T magnetic fields as shown in Fig. 3(d). The kinks are absent in Nd$_{1-x}$Sr$_x$NiO$_3$[37] and Sm$_{1-x}$Sr$_x$NiO$_3$[22]. It is possible that the abnormal magnetization in La$_{1-x}$Sr$_x$NiO$_3$ at 22 K origins from impurities during the synthesis process.

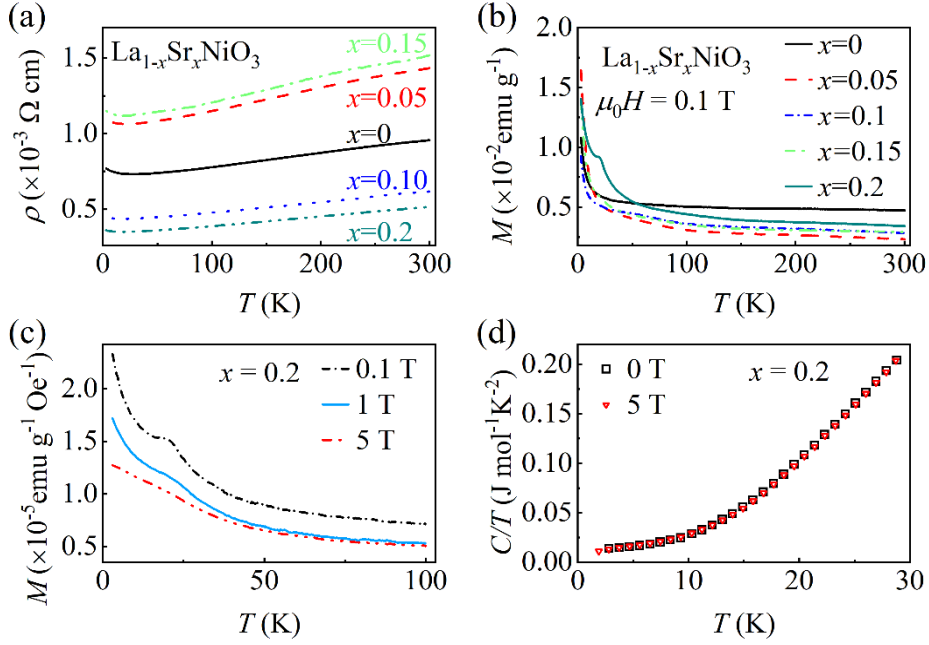

**Figure 3.** (a) Resistivity versus temperature $\rho(T)$ plots of La$_{1-x}$Sr$_x$NiO$_3$ ($0 \leq x \leq 0.2$). (b) Magnetic susceptibility of La$_{1-x}$Sr$_x$NiO$_3$ ($0 \leq x \leq 0.2$) with a field of $\mu_0 H = 0.1$ T. (c) Magnetization for La$_{0.8}$Sr$_{0.2}$NiO$_3$ in ZFC process for magnetic fields of $\mu_0 H = 0$, 1, and 5 T. (d) Specific heat of La$_{0.8}$Sr$_{0.2}$NiO$_3$ for $3 \leq T \leq 30$ K, and $\mu_0 H = 0$ and 5 T.

To explore the possible superconductivity in the bulk samples of La$_{1-x}$Sr$_x$NiO$_2$, we performed resistivity and magnetization measurements of La$_{1-x}$Sr$_x$NiO$_2$ ($0 \leq x \leq 0.2$) as shown in Fig. 4. The resistivity versus temperature $\rho(T)$ of La$_{1-x}$Sr$_x$NiO$_2$ in zero applied magnetic field in Fig. 4(a) reveals semiconducting-like behavior. The doped Sr does not affect the electronic properties severally. No superconductivity is observed in the bulk samples. In Fig. 4(b), we show the fittings by using the activation-energy model $\rho(T) = \rho_0 \exp(E/k_B T)$ through fitting the high temperature resistivity data from 200-300 K, where $\rho_0$ is a prefactor and $k_B$ is the Boltzmann constant. The resultant thermal activation energies are 48.9 ($x = 0$), 55.3 ($x = 0.05$), 54.2 ($x = 0.10$), 51.0 ($x = 0.15$), and 48.9 meV ($x = 0.20$). To explore the magnetic properties of La$_{1-x}$Sr$_x$NiO$_2$, we show magnetization and magnetoresistance as functions of magnetic field and temperature in Fig. 4(c) and (d). The magnetization measured in field cooling (FC) deviates from that of ZFC in a wide temperature range, indicating a spin glass state. Figure 4 (d) displays temperature dependence of the magnetoresistance (MR) of LaNiO$_2$. The MR effect changes signs from negative at 3, 10, and 20 K to positive at 30 and 50 K, similar to the observation in films of Nd$_{0.8}$Sr$_{0.2}$NiO$_2$[23]. The reversal signs of the MR upon changing temperature suggest existence of magnetic order or magnetic correlations. Figure 4 (e) shows a typical hysteresis of a ferromagnetism vs magnetic fields at various temperatures measured at $T = 3$, 50, 300 and 650 K. The ferromagnetism is reminiscent of the resident nickel in Sr doped $R$NiO$_2$ [22, 23, 38]. Therefore, magnetic susceptibility is measured from 300 to 1000 K to verify the existence of Ni. As shown in Fig. 4(f), the ferromagnetic transition of Ni at $T_c = 627$ K is identified in the bulk LaNiO$_2$. However, XRD measurements on powder samples of

LaNiO$_2$ did not reveal the existence of Ni, indicating a low content of Ni impurity in the samples. Thus, the spin-glass like behavior may be due to the Ni impurity. Small amount of Ni in the sample would not change the electronic transport properties. The MR effect should be intrinsic properties of LaNiO$_2$ and induced by magnetic correlations that have been proposed by Raman scattering measurements on NdNiO$_2$[39] and X-ray magnetic linear dichroism measurements on Nd$_{0.8}$Sr$_{0.8}$NiO$_2$ films[40].

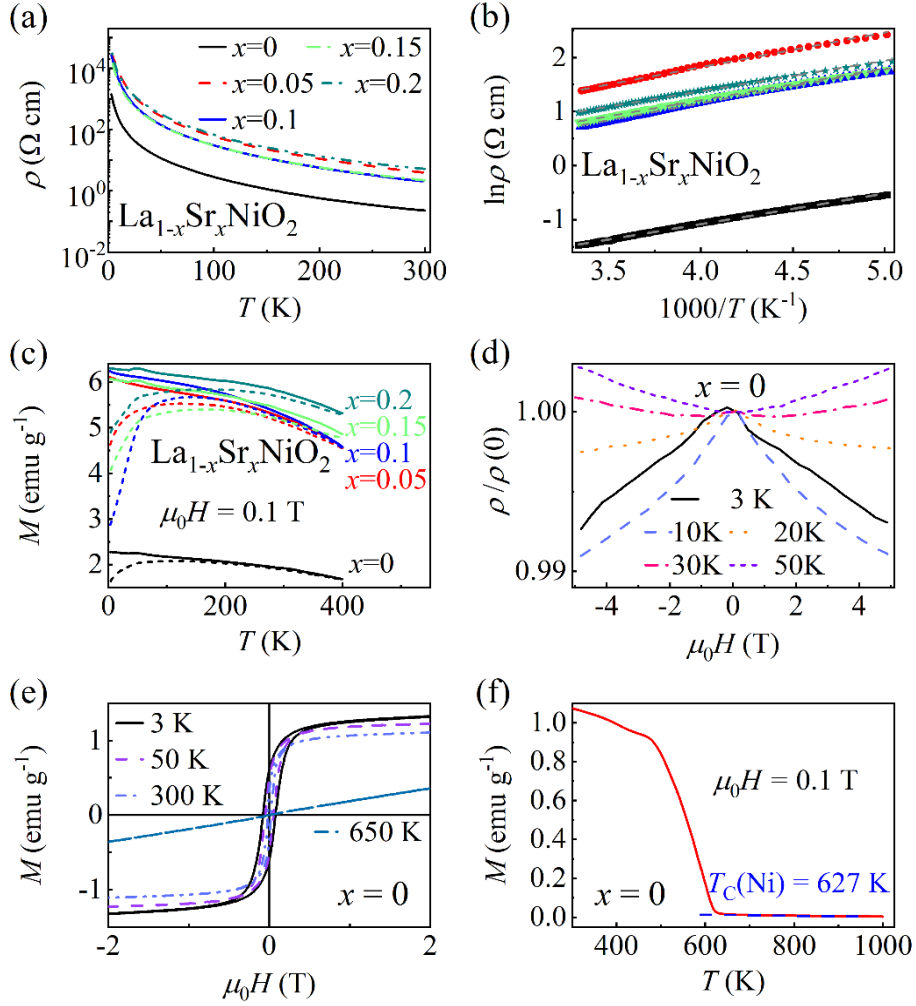

**Figure 4.** (a) Resistivity of La$_{1-x}$Sr$_x$NiO$_2$ ($0 \leq x \leq 0.2$) as a function of temperature from 3 to 300 K and (b) fits using the activation-energy model $\rho(T) = \rho_0 \exp(E/k_B T)$ from 200 to 300 K. (c) Magnetic susceptibility measured under ZFC and FC conditions for $3 \leq T \leq 400$ K and $\mu_0 H = 0.1$ T. (d) Magnetoresistance for LaNiO$_2$ at $T = 3, 10, 20, 30$, and 50 K. (e) Magnetization of LaNiO$_2$ at $T = 3, 50, 300$, and 650 K. (f) Magnetic susceptibility of LaNiO$_2$ from 300 to 1000 K with $\mu_0 H = 0.1$ T.

As theoretical calculations, $R$NiO$_2$ are metal[8, 46-48]. However, resistivity measurements reveal that bulk samples of $R$NiO$_2$ are insulating. It has been suggested that the difference between theoretical predictions and experimental observations could be ascribed to hydrogens that take place the apical oxygen sites to reduce the thermodynamical instability[49]. Small size of hydrogen can insert into the samples and form oxide hydride La$_{1-x}$Sr$_x$NiO$_y$H$_z$ which exhibit insulating behavior. This has been observed in NdNiO$_x$H$_y$ ($x\sim$2.3, $y\sim$0.7)[50], BaTiO$_{3-x}$H$_x$[51], Sr$_3$Co$_2$O$_{4.33}$H$_{0.84}$[52], and LaSrCoO$_3$H$_{0.7}$[53]. In addition, impurity and incomplete reduction of the apical oxygens in $R$NiO$_2$ could also result in large resistivity[30, 54]. It is known that the purity of films is essential for realizing superconductivity in Nd$_{1-x}$Sr$_x$NiO$_2$ films[41, 42]. The absence of superconductivity in bulk samples may be due to the deviation of O$^{2-}$ ions amount and the existence of Ni impurity. The interfaces between the substrate of SrTiO$_3$ and La$_{1-x}$Sr$_x$NiO$_2$ films may play a role in the magnetic correlations and emergence of superconductivity[29, 43-45]. Further studies on high quality single crystals of La$_{1-x}$Sr$_x$NiO$_2$ are required.

In conclusion, we synthesized polycrystalline samples La$_{1-x}$Sr$_x$NiO$_3$ and La$_{1-x}$Sr$_x$NiO$_2$ ($0 \leq x \leq 0.2$) and characterized their properties. The samples of La$_{1-x}$Sr$_x$NiO$_3$ are prepared using KOH as the flux. La$_{1-x}$Sr$_x$NiO$_2$ are synthesized by removing the apical oxygen from La$_{1-x}$Sr$_x$NiO$_3$ using CaH$_2$ as a reducing agent. La$_{1-x}$Sr$_x$NiO$_3$ are good metals as expectation. While La$_{1-x}$Sr$_x$NiO$_2$ exhibit insulating behavior, differing from theoretical calculations. No evidence of superconductivity is observed. The Ni impurity, resident apical oxygens, inserted hydrogens in bulk La$_{1-x}$Sr$_x$NiO$_2$ may mask the fundamental magnetic properties and inhibit the achievement of superconductivity.

**Acknowledgements**
Work at Sun Yat-Sen University was supported by the National Natural Science Foundation of China (Grants No. 12174454, 11904414, 11904416, U2130101), the Guangdong Basic and Applied Basic Research Foundation (No. 2021B1515120015), and National Key Research and Development Program of China (No. 2019YFA0705702).

Corresponding author: wangmeng5@mail.sysu.edu.cn